\documentclass[lettersize,journal]{IEEEtran}

\usepackage{balance}
\usepackage[utf8]{inputenc} 
\usepackage[T1]{fontenc} 

\usepackage{comment}

\usepackage{verbatim}
\usepackage{textcomp} 
\usepackage{gensymb} 

\def\BibTeX{{\rm B\kern-.05em{\sc i\kern-.025em b}\kern-.08em
    T\kern-.1667em\lower.7ex\hbox{E}\kern-.125emX}}
    
\usepackage{authblk} 

\usepackage{graphicx} 
\usepackage[caption=false,font=normalsize,labelfont=sf,textfont=sf]{subfig}
\usepackage{caption} 

\usepackage{amsmath, amsfonts}
\usepackage{algorithmic}

\usepackage{url} 

\usepackage{hyperref} 

\usepackage{lineno}
\usepackage{cleveref}
\usepackage[superscript]{cite}

\usepackage[nolist,nohyperlinks]{acronym}

\usepackage{array}
\usepackage{tabularx}

\usepackage{booktabs,siunitx}

\usepackage{multirow} 
\usepackage{multicol} 
\usepackage{makecell} 
\setcellgapes{4pt} 

\usepackage{colortbl}
\usepackage[table,dvipsnames,svgnames,x11names]{xcolor}
\usepackage{soul}

\usepackage{float} 
\usepackage{stfloats}

\usepackage
[a4paper,left=2.25cm,right=1.75cm,top=2cm,bottom=3cm]{geometry}

\begin{document}

\title{A Markov Chain Model for Identifying Changes in Daily Activity Patterns of People Living with Dementia} 

\author[1,3*]{Nan Fletcher-Lloyd}
\author[2,3]{Alina-Irina Serban}
\author[1,3]{Magdalena Kolanko}
\author[1,3]{David Wingfield}
\author[1,3]{Danielle Wilson}
\author[3,4]{Ramin Nilforooshan}
\author[1,3]{Payam Barnaghi}
\author[1,3]{Eyal Soreq}

\affil[1]{Department of Brain Sciences\\
        Imperial College London\\
        London, UK}
\affil[2]{Dyson School of Design Engineering\\
        Imperial College London\\
        London, UK}
\affil[3]{Care Research \& Technology Centre\\
        The UK Dementia Research Institute}
\affil[4]{Surrey and Borders Partnership NHS Foundation Trust}
\affil[*]{Corresponding Author}

\IEEEpubid{\begin{tabular}[t]{@{}c@{}}©2023 IEEE. Personal use of this material is permitted. Permission from IEEE must be obtained for all other uses, in any current or future media, \\including reprinting/republishing this material for advertising or promotional purposes, creating new collective works, for resale or redistribution to \\servers or lists, or reuse of any copyrighted component of this work in other works.\end{tabular}}

\maketitle

\begin{abstract}
Malnutrition and dehydration are strongly associated with increased cognitive and functional decline in people living with dementia (PLWD), as well as an increased rate of hospitalisations in comparison to their healthy counterparts. Extreme changes in eating and drinking behaviours can often lead to malnutrition and dehydration, accelerating the progression of cognitive and functional decline and resulting in a marked reduction in quality of life. Unfortunately, there are currently no established methods by which to objectively detect such changes. Here, we present the findings of an extensive quantitative analysis conducted on in-home monitoring data collected from 73 households of PLWD using Internet of Things technologies. The Coronavirus 2019 (COVID-19) pandemic has previously been shown to have dramatically altered the behavioural habits, particularly the eating and drinking habits, of PLWD. Using the COVID-19 pandemic as a natural experiment, we conducted linear mixed-effects modelling to examine changes in mean kitchen activity within a subset of 21 households of PLWD that were continuously monitored for 499 days. We report an observable increase in day-time kitchen activity and a significant decrease in night-time kitchen activity (t(147) = -2.90, p < 0.001). We further propose a novel analytical approach to detecting changes in behaviours of PLWD using Markov modelling applied to remote monitoring data as a proxy for behaviours that cannot be directly measured. Together, these results pave the way to introduce improvements into the monitoring of PLWD in naturalistic settings and for shifting from reactive to proactive care.\end{abstract}

\section{Introduction}
    \setlength{\parindent}{1cm}

\IEEEPARstart{T}{here} are currently around 50 million people living with dementia (PLWD) worldwide, and this number is estimated to rise to approximately 150 million by 2050 \cite{livingston2020dementia,PattersonADI2018}. Malnutrition and dehydration are strongly associated with increased cognitive and functional decline in PLWD \cite{Borda, HydroHypothesis, Lauriola, Nagae}. Furthermore, PLWD are more likely to experience malnutrition and dehydration when compared to age-matched controls \cite{Jesus2012,Siervo2014}, with such events accounting for ten times more hospital admissions in PLWD \cite{Natalwala2008}. Extreme changes in eating and drinking behaviours often result in episodes of malnutrition and dehydration. With malnutrition and dehydration being linked to the acceleration of cognitive and functional decline, this can lead to a marked reduction in quality of life. As such, it is crucial to detect changes in the in-home eating and drinking habits of PLWD. Despite this, there are currently no established methods by which to objectively detect such changes. 

There are many reasons why a change in eating and drinking habits might occur. The cognitive and functional decline symptomatic of a dementia diagnosis can increase the likelihood of physical difficulties in food preparation and feeding (having trouble chewing and swallowing), as well as shopping \cite{Keller2016,Roque2013}, and these changes tend to vary dependent on the progression of the disease \cite{Abdelhamid2016}. Another reason may be that the PLWD is experiencing a change in their state of health due to an underlying adverse health condition. Additionally, external factors that occur outside the disease model, such as a change in household occupancy or a catastrophic event, can also affect the eating and drinking behaviours of PLWD.

\IEEEpubidadjcol

In this study, we use in-home monitoring systems to detect changes in behavioural patterns of PLWD. The emergence of Internet of Things (IoT) technologies provides us with an unprecedented ability to continuously monitor and quantify behavioural patterns of PLWD passively and in real-world settings \cite{Ahamed2020,Kelly2020}. Recently, there have been several studies that have focused on the benefits of assistive living technologies for PLWD. For instance, Ishii \textit{et al.} and Urwyler \textit{et al.} employed passive infrared motion (PIR) sensors to monitor activities of daily living to differentiate between healthy controls and PLWD. By monitoring the continuity and regularity of performance of these activities, these studies were able to suggest a method by which to identify the onset of neurodegenerative dementias in the earlier stages of disease \cite{Ishii2016, Urwyler2017}. Previous work has also been done on time series anomaly detection applied to in-home monitoring data. For instance, Monekosso \textit{et al.} employed the Hidden Markov Model (HMM) technique to model behaviour from sensor data that had undergone distance-based clustering to group the activities of daily living of a PLWD \cite{Monekosso}. Despite being one of the most widely used techniques, HMMs suffer two main limitations: they do not take into account the temporal patterns generated from sensors, and their Black Box approach leads to reduced explainability of the model's decision-making process. As such, the focus of our current work is on developing a 'Glass Box' approach to anomaly detection that considers temporality.

Here, we use COVID-19 as a natural experiment to observe how household routines can be affected by external catastrophic events. From the start of the Coronavirus pandemic in December 2019 (COVID-19), governments worldwide have attempted to slow the spread of this disease using quarantining measures \cite{Rainero2020}. Since then, several reviews have reported changes in behavioural patterns at a population level \cite{Mignogna2021,Bennett2021}. These changes included increased snacking, as well as an increase in food intake (amount and frequency) \cite{Mignogna2021,Bennett2021}. Existing research has investigated the effects of lockdown on PLWD, with findings indicating that PLWD experienced changes in their in-home eating and drinking habits when in quarantine \cite{Cagnin2020,Rainero2020}. However, most of these studies used telephone questionnaires to collect this data \cite{Cagnin2020,Rainero2020}. Such an approach is not only overly reliant on self-reporting - which is not always feasible for PLWD - but could also be highly subjective. The use of remote monitoring technologies allows us to investigate previous statements in literature at a quantitative level. Further to this, we propose a novel 'Glass Box' analytical approach to detecting changes in the in-home behaviours of PLWD. 

\section{Methods}
    \setlength{\parindent}{1cm}

\subsection{Reproducibility}

To ensure this work is reproducible, we have made all the code involved in this work publicly available with a link included in this paper (see Section VI). Likewise, all datasets used in this work have been anonymized and also made publicly available with a link included in this paper (see Section VI). For all datasets shared, we include relevant statistics, a description of the pre-processing pipeline, and an explanation of any data that were excluded.  

\subsection{Participants}

Remote-monitoring data was collected from 120 households of PLWD as part of the ongoing Minder study being conducted at the UK Dementia Research Institute (UK DRI) Care Research \& Technology (CRT) centre \cite{crt, Minder}. This study uses activity within the kitchen as a proxy for eating and drinking habits. To be included in the Minder study, all individuals have to have an established diagnosis of dementia or mild cognitive impairment (see Supplementary Information). To be included in this study, each household had to have at least 100 days of kitchen activity data. Due to this criterion, a total of 73 households were included in the final cohort, with the mean number of days collected across these households being 427 \(\pm 257\), both to the nearest whole number. Of the 73 households, 20 PLWD were living alone (15 female and 5 male) and the remaining 53 households had multiple occupancy (17 female and 36 male) (see Figure \ref{fig:x fig1}a). Within this cohort, there were a range of diagnoses, with the majority of diagnoses being Alzheimer's Disease. For further information, see Supplementary Table 1. All participants were over the age of 50, with year of birth ranging from 1927 to 1962 and mean 1941 (to the nearest year) (see Supplementary Information Table 2 and Figure \ref{fig:x fig1}b)).

\begin{figure*}[h!]
\renewcommand{\thefigure}{1}
\setcounter{figure}{0}
\centering
\vspace{10mm}
    \includegraphics[width=\linewidth]{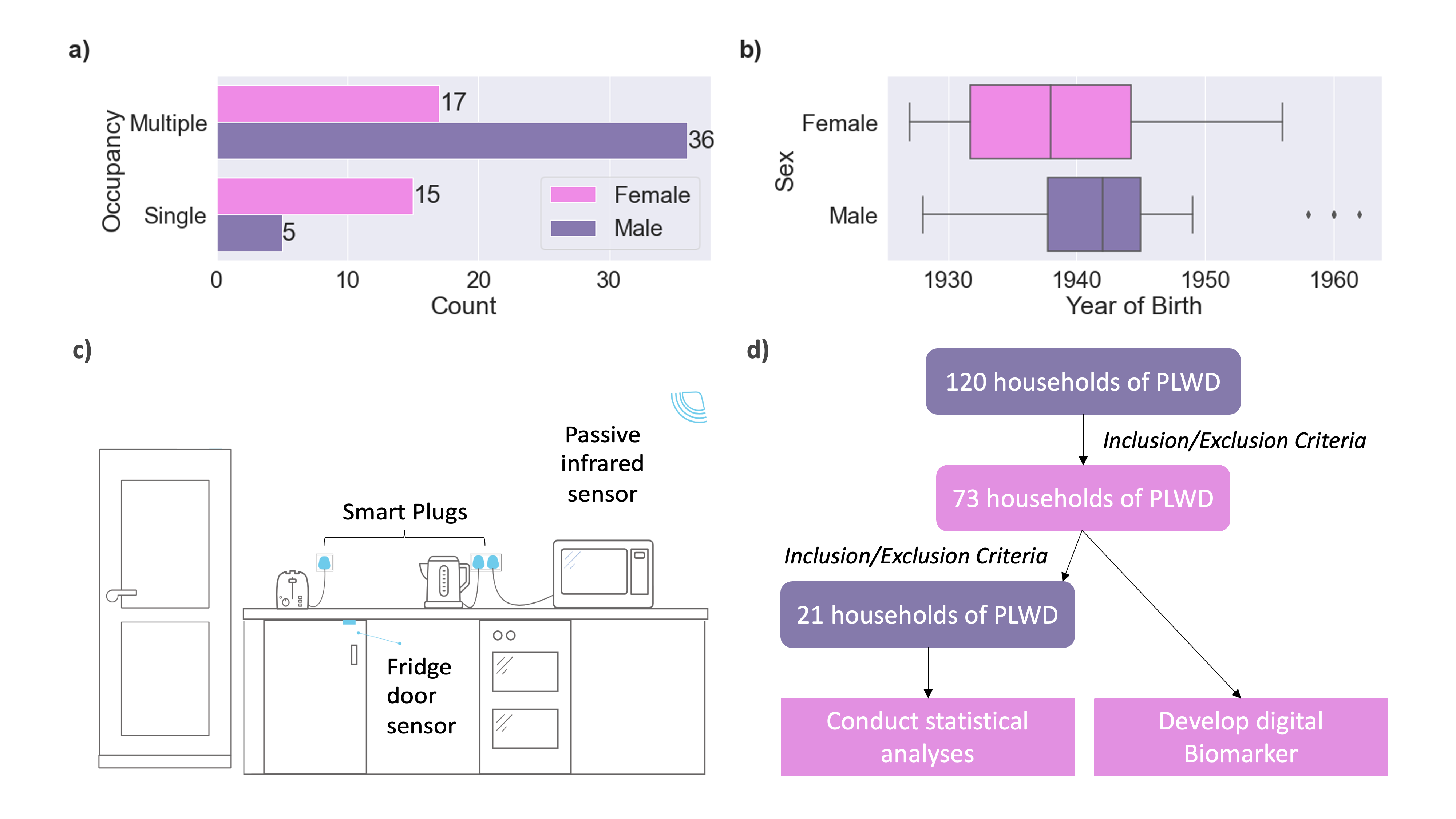}\hfil
\caption{Study cohort description, home monitoring system, and study design: a) \& b) Household occupancy, sex, and age distributions; c) sensors contributing to the data reported in this study: wall-mounted passive infrared motion sensors, a sensor on the fridge door, and smart plugs for the kitchen appliances (kettle and oven); and d) step-by-step workflow illustrating study design.}
\label{fig:x fig1}
\end{figure*}

\subsection{Information Governance and Ethics}

In this type of study, it is important to ensure that privacy and information governance requirements are fully considered and required approvals are obtained. The Minder study protocol received ethics approval from the London-Surrey Borders Research Ethics Committee then South West London Ethics Committee\footnote{ \href{https://www.hra.nhs.uk/about-us/committees-and-services/res-and-recs/search-research-ethics-committees/london-surrey-borders/}{https://www.hra.nhs.uk/about-us/committees-and-services/res-and-recs/search-research-ethics-committees/london-surrey-borders/}}. The information governance procedures have also been reviewed and approved by the Surrey and Borders NHS Trust information governance committee\footnote{\href{https://www.sabp.nhs.uk/aboutus/policies/digital-governance/infoGovPol}{https://www.sabp.nhs.uk/aboutus/policies/digital-governance/infoGovPol}} and the Imperial College information governance team\footnote{\href{https://www.imperial.ac.uk/admin-services/secretariat/college-governance/charters/policies-regulations-and-codes-of-practice/policy-framework/}{https://www.imperial.ac.uk/admin-services/secretariat/college-governance/charters/policies-regulations-and-codes-of-practice/policy-framework/}}. We obtained informed written consent from all the study participants. Each participant was assessed according to Mental Capacity Act guidelines\footnote{ \href{https://www.gov.uk/government/publications/mental-capacity-act-code-of-practice}{https://www.gov.uk/government/publications/mental-capacity-act-code-of-practice}} by a fully qualified researcher who has completed a mandatory clinical practice course. Participants understood the study and were able to understand the consent process.

\subsection{In-home Monitoring Technologies}

A range of Develco IoT devices\footnote{\href{https://www.develcoproducts.com/}{https://www.develcoproducts.com/}}, deployed by Howz\footnote{\href{https://www.howz.com/}{https://www.howz.com/}}, were used to collect anonymized binary data from the households of participants. All of the sensors were Develco products. Movement data was collected using a wall-mounted PIR Develco Mini Motion Sensor, information on the opening and closing of the fridge door collected using the Delveco Window Sensor, and Develco Smart Plug Minis collected information on the usage of appliances (kettle and oven) (see Figure \ref{fig:x fig1}c). Exact placement of these sensors varied respectively with the layout of individual households.

\subsection{Experiments}

In this section, we first describe a quantitative analysis on a subset of households of PLWD conducted to investigate the effect of the COVID-19 pandemic on the kitchen activities of PLWD. We then report on the development of an analytical approach by which to detect changes in the in-home eating and drinking habits of PLWD using kitchen activities as a proxy for eating and drinking behaviours.\\

\noindent \textbf{Investigating the Effect of the COVID-19 Pandemic on In-home Eating and Drinking Habits}: we report on the acquisition and treatment of data used in this analysis. We then describe the statistical tests conducted on the processed data by which to quantify the effect of the COVID-19 pandemic on the in-home eating and drinking habits of PLWD.\\

\noindent \textit{Data Acquisition}: the COVID-19 pandemic was used as a natural experiment, with data being continuously collected from 21 households of PLWD over a 499-day period from the 1st of December 2019 to the 12th of April 2021, a period inclusive of the three national UK lockdowns \cite{UK_COVID_Timeline}. Of the 21 households, 5 PLWD were living alone (3 female and 2 male) and the remaining 16 households had more than one occupant (6 female and 10 male). In this subset of participants, there was an age range of 72 to 92 years old and mean age 80.0 years old. All households had a PIR motion sensor, a fridge door sensor, a smart plug for the kettle, and at least one smart plug for a toaster, microwave, or oven appliance.

This study analyses kitchen activity levels at seven progressive periods across a 17-month time-span. Each period represents a specific stage of the COVID-19 pandemic in relation to the appearance of COVID-19 in the UK and the successive measures put in place by the UK government in response. A pre-COVID baseline was derived from data collected prior to the first officially recorded infections in the UK (P1). Period two (P2) defines the time from the initial onset of COVID-19 in the UK up to the first lockdown. The third, fifth, and seventh periods (P3, P5, and P7) denote the first, second, and third UK lockdowns, respectively. The relaxation period between the first and second lockdowns is denoted by period four (P4), while the period of relaxation with continued tier 4 level restrictions between the second and third lockdowns is denoted by period six (P6). The complete timeline is as follows:
\vspace{1em}
\begin{itemize}
    \item P1 - pre-COVID baseline (01/12/2019 - 30/01/2020)
    \item P2 - the onset of COVID-19 in the UK (31/01/2020 - 23/03/2020)
    \item P3 - first UK lockdown (24/03/2020 - 01/06/2020)
    \item P4 - relaxation (02/06/2020 - 05/11/2020)
    \item P5 - second UK lockdown (06/11/2020 - 02/12/2020)
    \item P6 - end of second lockdown but with continued restrictions (03/12/2020 - 06/01/2021)
    \item P7 - third UK lockdown (07/01/2021 - 12/04/2021)
\end{itemize}

All households had missing data between the dates of 31/07/2020 and 03/08/2020 inclusive, due to a network failure.\\

\noindent \textit{Data Pre-Processing}: in our households, activity is triggered and logged sparsely per event, with seconds precision and a 30 second delay. For our analyses, we use an aggregate measure of kitchen activity. Mean household kitchen activity is calculated based on the sum of mean daily activity across the different kitchen sensors (see Figure \ref{fig:x fig3}).

Our second analysis focused on kitchen activity at different times of the day. For this, sensor activity is re-sampled into four six-hourly periods (three daytime periods: morning (06:00 - 12:00); afternoon (12:00 - 18:00); and evening (18:00 - 00:00), and one night-time period (00:00 - 06:00)). Mean kitchen activity at each time of day is then calculated based on the sum of mean activity for that time of day across the different kitchen sensors (see Figure \ref{fig:x fig4}). To remove interference from what we already knew would be a significant effect of time of day on kitchen activity levels, the mean kitchen activity of each household is standardized across all households by time of day.\\

\noindent \textit{Statistics}: linear mixed-effects (LME) modelling, using the R packages (lme4/lmerTest) in RStudio \cite{lmertest}, was used to test the relationship between the onset and progression of the COVID-19 pandemic in the UK and kitchen activity. A linear mixed model can be specified in matrix form as:

\begin{equation}
\label{eq1}
y = \displaystyle X\beta + Zu + \epsilon
\end{equation}

with y representing the outcome variable (in this case, kitchen activity), \mbox{$\beta$} representing all fixed-effects parameters, u the random effects, X the n x p design matrix for the fixed-effects parameters, Z the n x q design matrix for the random effects, and \mbox{$\epsilon$} the residual effects.
We modelled the fixed effects of the pandemic periods, home occupancy, and time of day alongside the random effects associated with household heterogeneity. Unless otherwise stated, we use standard significance reporting (p < 0.0001 denoted as ***, p < 0.001 denoted as **, p < 0.05 denoted as *, and p < 0.1 denoted as $\cdot$).\\

\noindent \textbf{Developing a Markov Chain Model}: we define an anomaly as changes in behaviour that significantly deviate from stable, predictable behavioural patterns. However, in order to define an anomaly as such, we must first be able to quantify regular behavioural patterns. We wanted to discover patterns in the sequences of sensor recordings in such a way as to be able to describe anomalous behavioural patterns as significant deviations from stable, predictable behaviour. The simplest Markov model is a Markov Chain. A Markov chain describes a sequence of possible transitions or changes of state of a system to which are assigned a probability dependent only on the state attained in the previous transition. Stochastic matrices are square matrices describing these transitions and can be used as behavioural patterns reflecting in-home activities. From here on, we will refer to our stochastic matrices as transition matrices.

In this study, due to the sequence of possible transitions always occurring in the kitchen, the number of sensor firings recorded by the PIR kitchen motion sensor is very high, with kitchen-to-kitchen transitions being the most frequently occurring transition. As such, to better control the level of noise in the data, we chose to eliminate kitchen-to-kitchen transitions from our transition matrices. For the same reason, as the fridge door sensor recorded both opened-to-closed and closed-to-opened transitions, and as the latter transition was extremely likely to occur sequentially from the former, we only included opened-to-closed transitions. For any one transition matrix, we have 16 transitions (see Figure \ref{fig:x fig2}): kitchen-to-kitchen (which was always set to 0), kitchen-to-kettle, kitchen-to-fridge, kitchen-to-oven, kettle-to-kitchen, kettle-to-kettle, kettle-to-fridge, kettle-to-oven, fridge-to-kitchen, fridge-to-kettle, fridge-to-fridge, fridge-to-oven, oven-to-kitchen, oven-to-kettle, oven-to-fridge, and oven-to-oven. Our transition matrices are right stochastic matrices, meaning that the probabilities in each row sum to 1.\\

\noindent \textit{Algorithm Design}: to extract transition matrices across any period, we use a sliding window with a pre-set step frequency (see Figure \ref{fig:x fig2}). For any point in time as defined by the step frequency of the sliding window, these matrices can be compared across the current window and that of the baseline window, both of which could be altered to allow for the most clinically relevant windows of time to be examined (see Figure \ref{fig:x fig2}). Each transition probability is derived from the sum of that transition divided by the sum of the total number of transitions from that state, at a set re-sampling rate, aggregated across both the current and the baseline window. 

The duration of the current window, baseline window, and step, as well as the re-sampling frequency of the transition matrices were curated by a trial-and-error approach in which to select for smoothing that reduced the effect of noise and highlighted important trends in the dataset. Such an approach also allows us to mitigate the effect of missing data. \\

\noindent \textit{Measuring Similarity}: we use the Frobenius distance to calculate a measure of similarity between any two matrices (see Figure \ref{fig:x fig2}). The Frobenius distance is the square-root of the sum of the squared distances between each of the singular values of one matrix and their respective values in another matrix (see Eq. \ref{eq2}). In this instance, the higher the Frobenius distance, the greater the dissimilarity between any two matrices, with the measure capped at 4 for each matrix (the closer to 0, the more similar, and the closer to 4, the more dissimilar).

\begin{equation}
\label{eq2}
F_D = \displaystyle \sqrt{\sum_{i,j} (a_{ij} - b_{ij})^2}
\end{equation}

\begin{figure*}[t]
\renewcommand{\thefigure}{2}
\setcounter{figure}{0}
\centering
\vspace{10mm}
    \includegraphics[width=\linewidth]{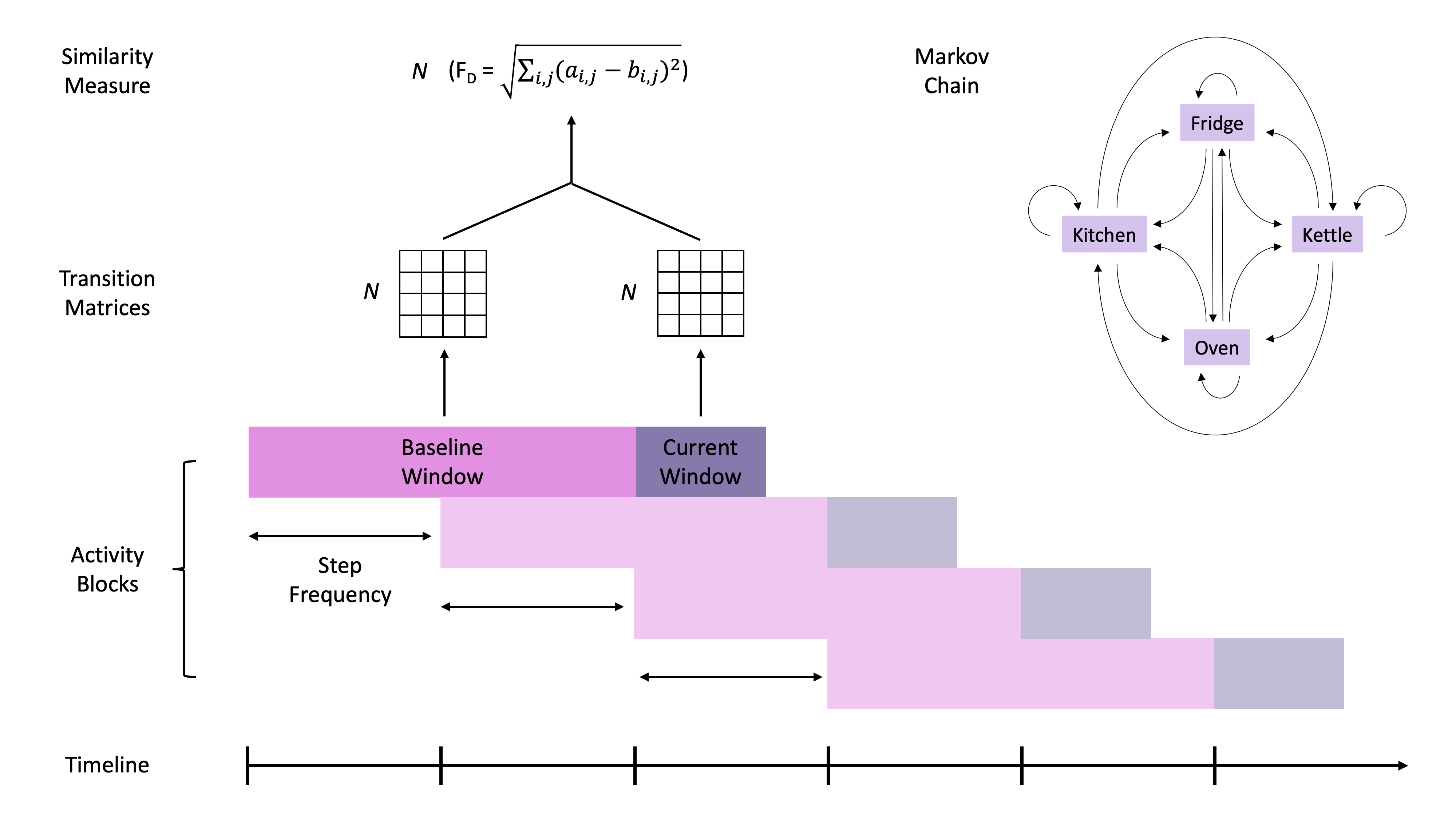}\hfil
\caption{Diagram of sliding window algorithm used to extract dissimilarity measures.}
\label{fig:x fig2}
\end{figure*}

\noindent \textit{{Proof-of-concept Study}}: to provide proof-of-concept for this approach, we employed a retrospective analysis using real-world data across 73 households of PLWD using a case-by-case approach. For this study, we set the step frequency of our sliding window to 1 day and the current and baseline windows to 1 and 3 weeks, respectively. Initial transition matrices were derived using an hourly re-sampling rate. Time windows were chosen based on clinical relevance, balancing the need to detect both episodic and gradual changes. 
\section{Results}
    \setlength{\parindent}{1cm}

\subsection{Higher Activity Levels Within the Kitchen During the Pandemic}

Data was collected from a subset of 21 households of PLWD between the 1st of December 2019 and the 12th of April 2021 using a passive infrared motion sensor, a door sensor on the fridge door, and smart plugs for the kitchen appliances (see Figure \ref{fig:x fig1}c). In total, over 2 million unique observations were recorded across 499 days. Changes in activity patterns were examined over seven time periods: P1 - acted as a baseline for kitchen activity (from the 1st of December 2019 to the 30th of January 2020); P2 - the onset of COVID-19 in the UK with the first infections being recorded (the 31st of January 2020 to the 23rd of March 2020); P3 - the introduction of the first UK lockdown with a stay-at-home ruling announced alongside measures such as social distancing and self-isolation (the 24th of March 2020 to the 1st of June 2020); P4 - relaxation of the first UK lockdown and introduction of restrictions focused on local lockdowns and remote working (the 2nd of June 2020 to the 5th of November 2020); P5 - the second UK lockdown (the 6th of November 2020 to the 2nd of December 2020); P6 - relaxation of the second UK lockdown but with continued Tier 4 level restrictions (the 3rd of December 2020 to the 6th of January 2021); and P7 - the third UK lockdown (the 7th of January 2021 to the 12th of April 2021).

Mean household kitchen activity was calculated based on the sum of mean daily activity across the different kitchen sensors. We used LME modelling and analysis of variance (ANOVA) to compare kitchen activity between the pre-COVID baseline (P1) and the pandemic periods (P2 - P7). Here, we report the t-statistics and the ANOVA F-statistics, with p-values, as measures of the difference between group means. All values are provided to three significant figures. The onset of the COVID-19 pandemic in the UK is associated with a significant increase in kitchen activity (F(1,21) = 27.9***, Figure \mbox{\ref{fig:x fig3}}a). The main effect of changes in activity due to COVID-19 can be seen in both single and multiple occupancy households (F(1,21) = 21.0***), but with no effect of occupancy (F(1,21) = 2.25, p=0.149) and no interaction between activity in the COVID-19 pandemic and occupancy (F(1,21) = 0.0257, p=0.874). A similar result was observed when the pandemic periods were modelled separately (see Figure \mbox{\ref{fig:x fig3}}b), with the main effect of COVID-19 being a significant increase in kitchen activity from P1 onwards (F(6,126) = 8.77***) but with no significant effect of occupancy(F(1,21) = 1.46, p = 0.240) and no interaction between the mean kitchen activity across the pandemic periods and household occupancy (F(6,126) = 0.622, p=0.712).

\begin{figure*}[h]
\renewcommand{\thefigure}{3}
\setcounter{figure}{0}
\centering
\vspace{10mm}
    \includegraphics[width=0.75\linewidth]{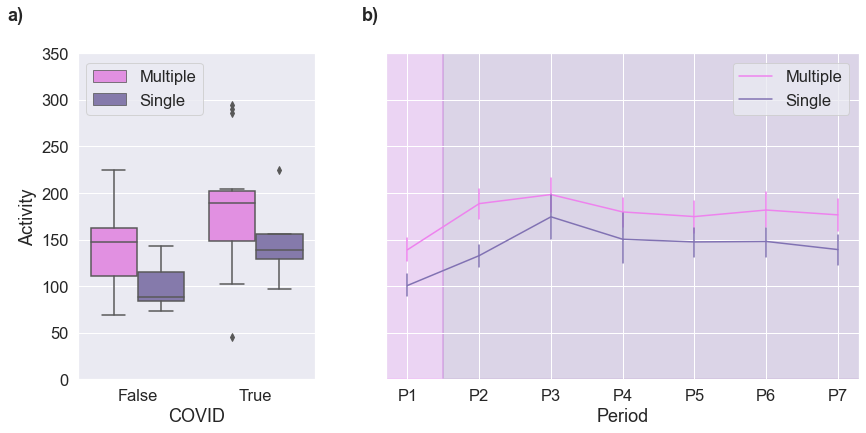}\hfil
\caption{Changes in mean kitchen activity during the pandemic according to household occupancy: a) mean kitchen activity pre-COVID and during COVID; and b) mean kitchen activity across all the pandemic periods, respectively.}
\label{fig:x fig3}
\end{figure*}

\begin{figure*}[h!]
\renewcommand{\thefigure}{4}
\setcounter{figure}{0}
\centering
\vspace{10mm}
    \includegraphics[width=0.85\linewidth]{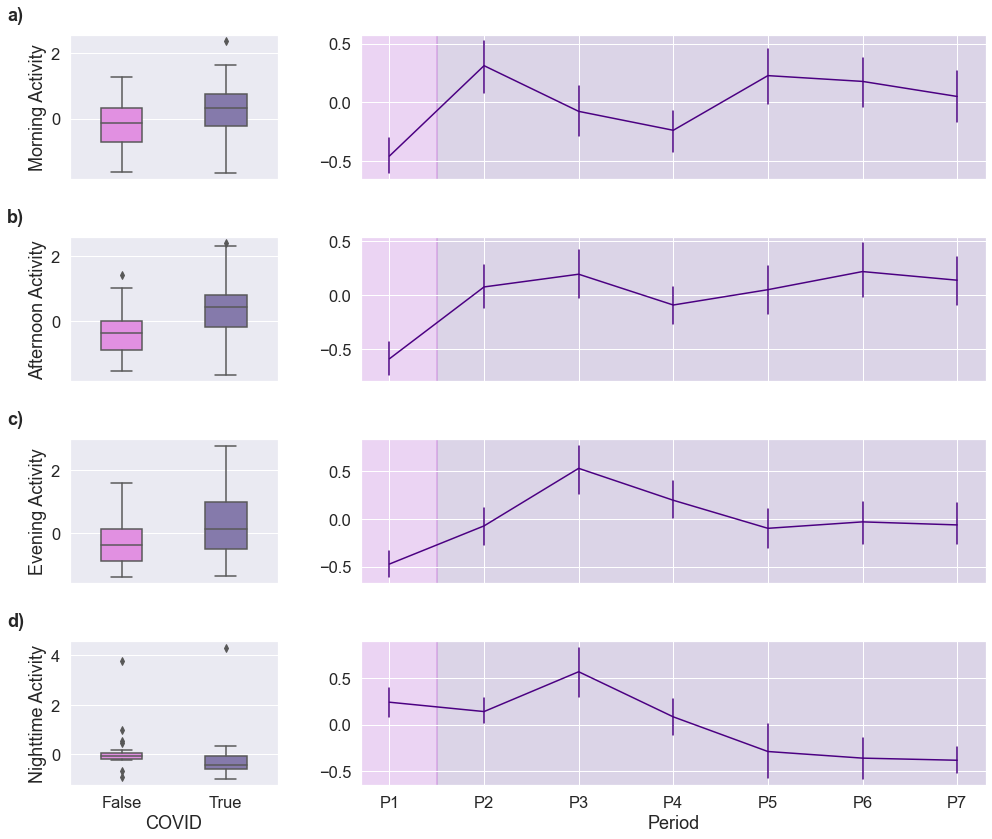}\hfil
\caption{Changes in mean kitchen activity during the pandemic during different times of the day: a-c) changes in kitchen activity during daytime hours (morning (06:00-12:00), afternoon (12:00-18:00), and evening (18:00-00:00); and d) change in kitchen activity during night-time hours (00:00-06:00).}
\label{fig:x fig4}
\end{figure*}
\vspace{1em}

The change in kitchen activity between the pre-COVID baseline (P1) and the pandemic periods (P2 - P7) was also specific to the time of day, with a significant interaction between COVID-19 and time of day (F(3,147) = 3.62*). As expected, all four times of the day exhibited a change in kitchen activity levels (see Figure \mbox{\ref{fig:x fig4}}a); however, only the nighttime hours saw a significant change in activity levels (t(147) = -2.90**). This interaction was also seen when the pandemic periods were modelled separately (F(18,567) = 2.74***). The daytime hours saw an initial increase in the level of kitchen activity during the earlier stages of the pandemic (P2 - P3) but activity began to return to baseline levels in the later stages. Night-time kitchen activity saw an initial increase (P3) followed by a continuous decrease (past baseline) from P4 to P7, without returning to baseline (see Figure \mbox{\ref{fig:x fig4}}b).

\subsection{Transition Matrices as Behavioural Patterns}

Having observed that a catastrophic event such as the COVID-19 pandemic significantly affects kitchen activity levels, we then set out to further investigate whether such changes are reflected in the daily behavioural patterns, with the aim being to develop an analytical approach that can detect these changes. 

Using kitchen activity data as a proxy, we derived transition matrices as behavioural patterns reflecting in-home eating and drinking habits throughout the day across time. We then used the Frobenius distance to gauge a measure of dissimilarity between subsequent transition matrices. We then conducted retrospective analyses on all 73 households of PLWD. 

Here, we present case studies from three households of PLWD (all multiple occupancy) as a proof-of-concept that our approach can be used to detect changes in behaviours. For each case study, the measure of dissimilarity is summed across each time of day (night, morning, afternoon, and evening) as initially introduced in Section 2.1, as we saw that there was a significant interaction between time of day at this re-sampling rate (6-hourly) and the trigger event (in the above case, the COVID-19 pandemic). The measure of dissimilarity for each time of day is capped at 4 x 6 hours = 24.\\ 

\noindent \textbf{Case Study 1}: we present the time series data from a home of a PLWD in our COVID cohort subset (see Figure \ref{fig:x fig5}). This study is also an exemplar of a household experiencing sustained changes in behaviour. For this household, transition matrices were extracted and compared during the 17-month period from the 1st of December 2019 to the 12th of April 2021, inclusive of the three UK lockdowns. As can be observed in Figure \ref{fig:x fig5}, the onset of each new period (as indicated on the figure) is accompanied by an immediate increase in dissimilarity across all times of day, suggesting a change in the patterns of behaviour in the kitchen. Particularly for P2 - P4, following the initial dramatic increases around the beginning of each period, dissimilarity begins to decrease as time progresses before once again increasing at the onset of a new period (see Figure \ref{fig:x fig5}). These results suggest that the change in kitchen behaviours was sustained for each period, respectively, with the household adapting to the changing situation induced by the severity of restrictions accompanying each period.\\ 

\begin{figure*}[h]
\renewcommand{\thefigure}{5}
\setcounter{figure}{0}
\centering
\vspace{10mm}
    \includegraphics[width=0.95\linewidth]{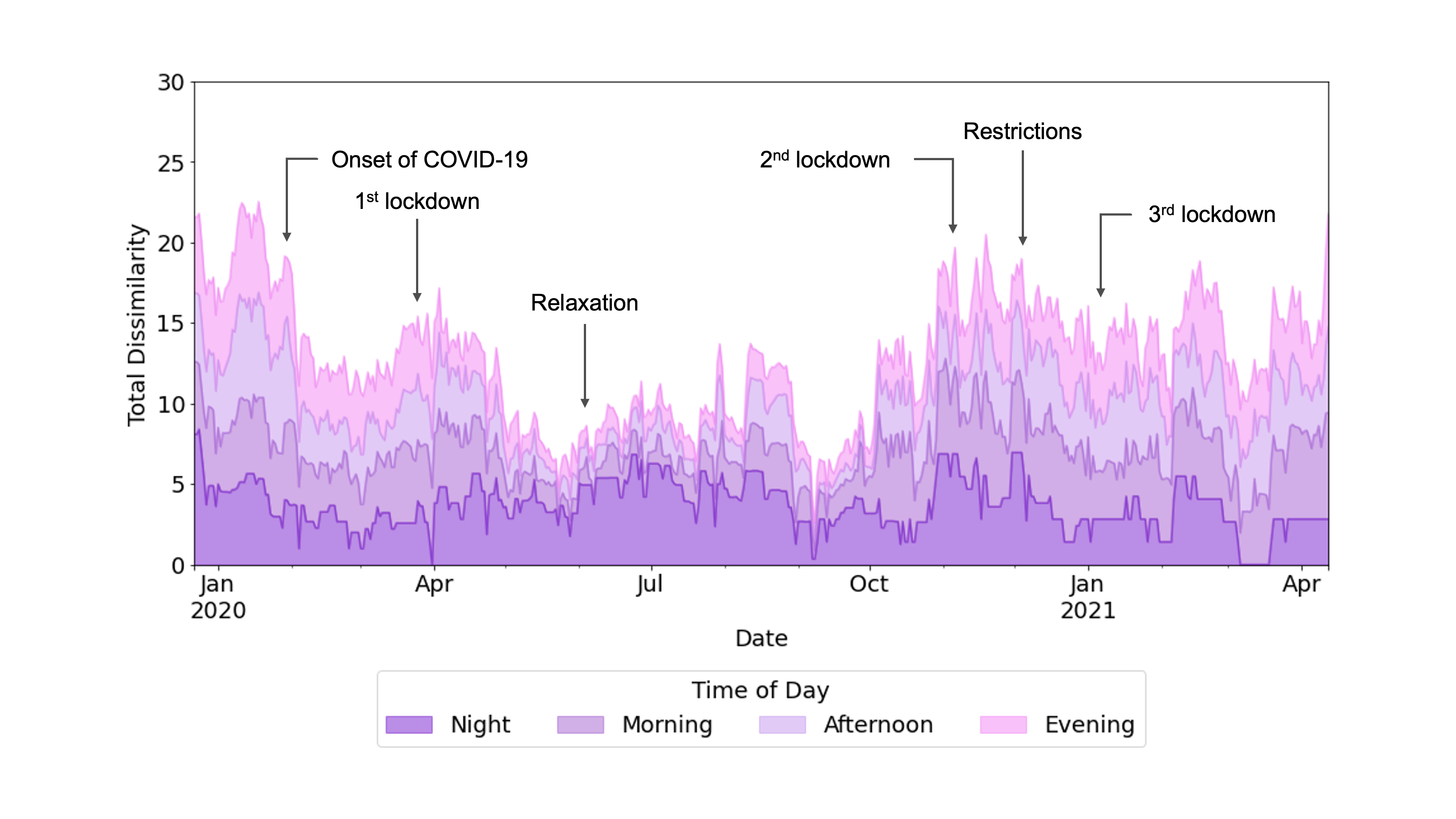}\hfil
\caption{Case study 1 time series data illustrating the output of the sliding window algorithm from a household of a PLWD in the COVID cohort subset. Total dissimilarity is summed across each 6-hour period of the day (night, morning, afternoon, and evening).}
\label{fig:x fig5}
\end{figure*}

\begin{figure*}[h!]
\renewcommand{\thefigure}{6}
\setcounter{figure}{0}
\centering
\vspace{10mm}
    \includegraphics[width=0.95\linewidth]{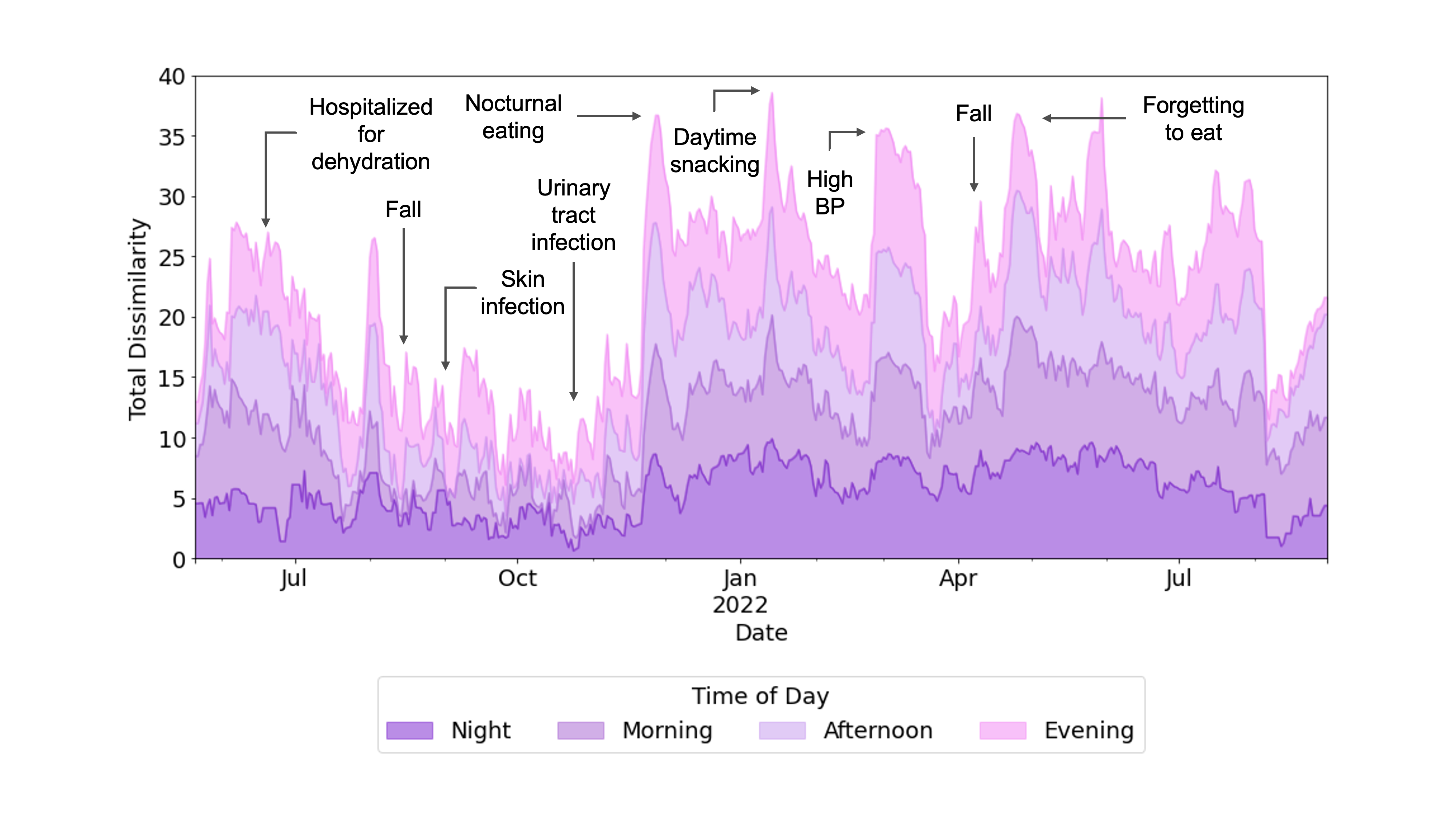}\hfil
\caption{Case study 2 time series data illustrating the output of the sliding window algorithm from a household of a PLWD who experienced several succesive carer-verified changes in their eating and drinking habits. Total dissimilarity is summed across each 6-hour period of the day (night, morning, afternoon, and evening).}
\label{fig:x fig6}
\end{figure*}

\begin{figure*}[h!]
\renewcommand{\thefigure}{7}
\setcounter{figure}{0}
\centering
\vspace{10mm}
    \includegraphics[width=\linewidth]{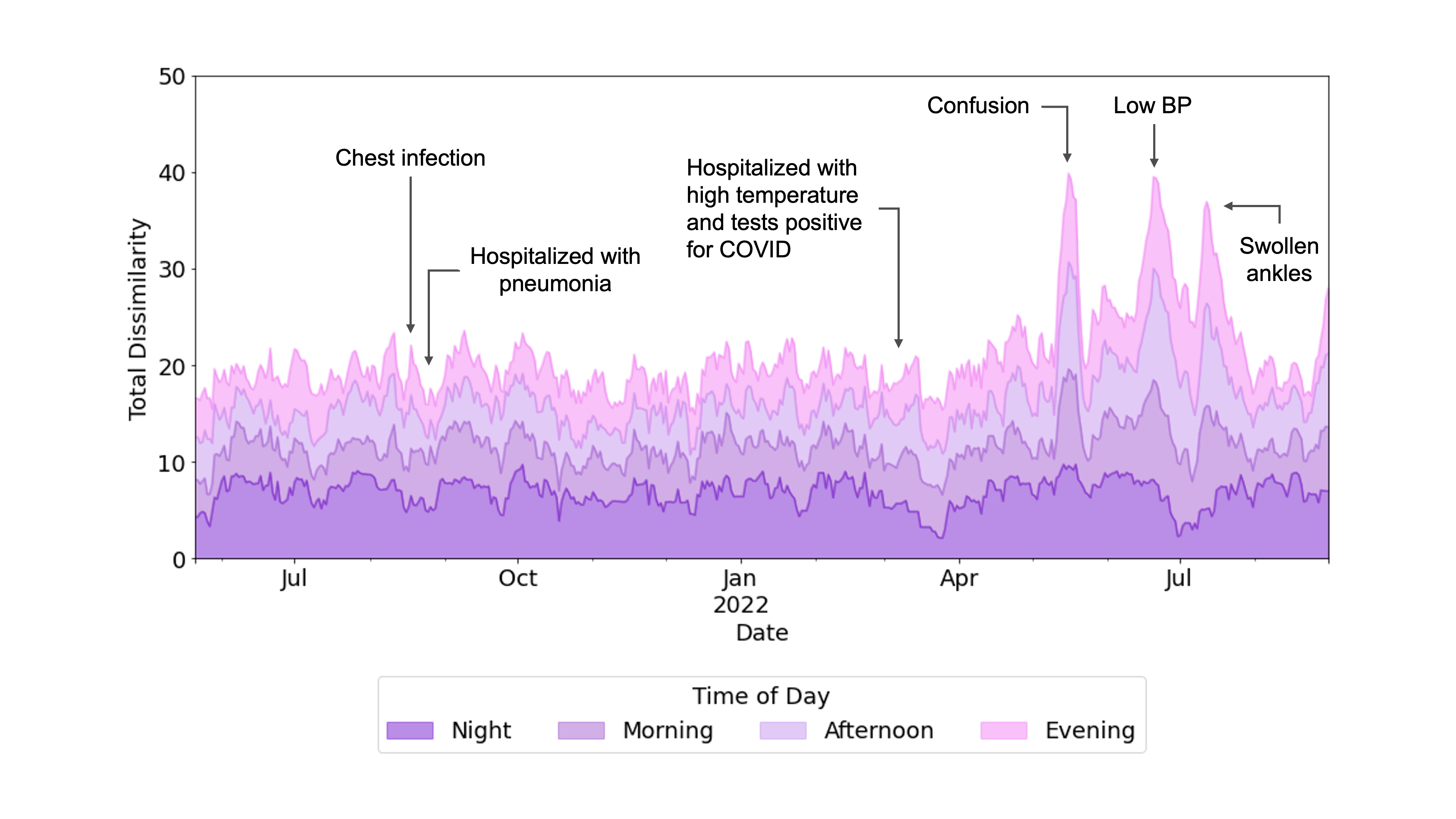}
\caption{Case study 3 time series data illustrating the output of the sliding window algorithm from a household of a PLWD who experienced adverse health events following a COVID diagnosis. Total dissimilarity is summed across each 6-hour period of the day (night, morning, afternoon, and evening).}
\label{fig:x fig7}
\end{figure*}

\noindent \textbf{Case Study 2}: the second case study focuses on a home of a PLWD who experienced several successive carer-verified changes in their eating and drinking habits. As such, this case study is an exemplar of a household experiencing cumulative changes. Transition matrices were extracted and compared from the 13th of April 2021 onwards (post-COVID-19 pandemic). As can be seen in Figure \ref{fig:x fig6}, dissimilarity in the kitchen behavioural patterns increased in this household in June 2021, with the PLWD being hospitalized mid-June due to dehydration. Our algorithm suggests a change in behaviour before this date, and the carer of this participant later confirmed that the PLWD had experienced a change in their eating and drinking habits prior to hospitalization. From the beginning of August to the end of November 2021, the PLWD experienced several health events in the form of infections and was prescribed several courses of antibiotics. Our results show dissimilarity is at its lowest around this period of time. As antibiotics usually have to be taken on an empty stomach (1 hour before eating or 2 hours after), we hypothesize that this might have led to more consistent eating and drinking habits. Late November, the participant's carer reported that the PLWD had been struggling to sleep, snacking during the night. This was then followed by reports of daytime snacking and an increased appetite in the middle of January 2022. As seen in Figure \ref{fig:x fig6}, around the dates these changes in eating behaviours were observed, there is a rapid increase in dissimilarity in the kitchen behavioural patterns across all four periods of the day. As we saw in case study 1, dissimilarity then begins to decrease, which we interpret as the new behaviours being sustained. This aligns with the PLWD experiencing continued episodes of night-time snacking and increased appetite, respectively. Across March 2022, the participant experienced several episodes of high blood pressure, which we interpret as the prolonged period of dissimilarity seen across this month (see Figure \ref{fig:x fig6}). Finally, the PLWD began to experience episodes of forgetfulness, particularly as regards to mealtimes. This was reported around late April 2022 and aligns with the sporadic changes in dissimilarity resulting from our algorithm from this point on (see Figure \ref{fig:x fig6}).\\ 

\noindent \textbf{Case Study 3}: finally, we present the time series data of a home of a PLWD who experienced several adverse health events following a COVID diagnosis in the first week of March 2022. As such, this case study provides an exemplar of a household experiencing episodic changes. Again, transition matrices were extracted and compared post the COVID-19 pandemic. This participant regularly experienced episodes of agitation accompanied by night-time wandering and low mood accompanied by restlessness during the day. This aligns with the relatively high levels of dissimilarity, particularly at night-time, when compared to the two previous case studies, that we observe in Figure \ref{fig:x fig7}. The PLWD had previously only experienced one health event when they were briefly hospitalized after developing pneumonia in August 2021 (see Figure \ref{fig:x fig7}). Around the time of their COVID diagnosis, the PLWD was reported to have experienced difficulty in moving around, which aligns with the decrease in dissimilarity shortly after this time (see Figure \ref{fig:x fig7}). We hypothesize that as the PLWD became less mobile, their wandering and restlessness reduced, leading to more consistent kitchen behaviours which we believe to be reflecting the routine of the carer. Following on from this event, the participant experienced a period of confusion with loss of time mid-May, a period of low blood pressure mid-June, and some further mobility issues due to swollen ankles from mid-July. Correspondingly, we see three narrow peaks indicating dramatic increases in the dissimilarity of kitchen behavioural patterns around these dates (see Figure \ref{fig:x fig7}).

\section{Discussion}
    \setlength{\parindent}{1cm}

We used passive in-home monitoring data to conduct retrospective analyses on 73 households of PLWD. Over 5 million unique observations were collected from the 1st of December 2020 to the 31st of August 2022, providing a unique opportunity to monitor the behavioural patterns of this vulnerable population over an extended period of time.

We first investigated the effect of the COVID-19 pandemic on the in-home kitchen activity of 21 households of PLWD, from which nearly 2 million unique observations were collected across nearly 500 days. In-home kitchen activity levels increased observably during daytime hours (morning (06:00 - 12:00), afternoon (12:00 - 18:00), and evening (18:00 - 00:00)) but decreased significantly during night-time hours (00:00 - 06:00) across households following a declaration from the World Health Organisation that COVID-19 was a Public Health Emergency of International Concern. This change preceded the introduction of the first stay-at-home ruling in the UK, suggesting that PLWD proactively changed their behaviour prior to the onset of public health restrictions. 

Investigating the effects of changes such as those that occurred due to the implementation of COVID-19 related restrictions is important for several reasons. PLWD are among the most clinically vulnerable patient group in the population. They have a higher rate of hospitalization in comparison to age-matched healthy controls, particularly for preventable adverse health conditions such as malnutrition and dehydration \cite{Phelan2012}. Yet, they have a reduced tendency to consult with healthcare professionals \cite{Cooper2017}. This work allows us to understand how well vulnerable populations, such as PLWD, adapt to both internal and external changes. 

Using COVID-19 as a natural experiment, this study provided a way to measure the effects of such an event on household activity at a quantitative level. Here, we provide a descriptive analysis by which to better understand previous statements in literature as regards the effects of pandemic quarantining on the in-home eating and drinking habits of PLWD. The strong correspondence between changes in kitchen activity and the onset of public health measures in the UK illustrates a distinct change in kitchen-related behaviours. However, it should be noted that our quantitative measures are based on proxy kitchen movement and appliance use activities. While these observation and measurement data do not reflect the complete picture of in-home eating and drinking habits, the continuous and relatively long period of data collection does provide a unique opportunity to analyse the changes in patterns of related activities. 

In this paper, we also present a novel analytical approach by which we might proactively detect changes in the behaviours of PLWD. We derived transition probabilities of a Markov chain of kitchen activity as behavioural patterns reflecting in-home eating and drinking habits throughout the day and used a dissimilarity measure to compare between subsequent transition matrices across time. Initial findings from the case studies presented show that, when applied to real-world data, our approach can detect sustained, cumulative, and episodic changes in behaviour.

By designing an algorithm that can be manually tuned to search through clinically relevant windows, we have laid the foundation for a 'Glass Box' approach to anomaly detection that is generalizable across any cohort. Furthermore, this method also allows anomaly detection to be patient specific. Our algorithm was applied to each household individually, meaning we are able to derive unique behavioural profiles that allowed us to quantitatively define stable behavioural patterns as those that lay within the range of natural variation in dissimilarity for any household. 

\section{Conclusions and Future Directions}
    To summarize, the results of this study further demonstrate the utility of remote data collection using IoT technologies, showing that remote monitoring data can be an effective proxy by which to study behaviours that cannot be directly measured, while allowing PLWD to remain in the comfort of their own homes. Our work expanded on the capabilities of in-home monitoring devices to identify changes in the behavioural patterns of PLWD and provide quantifiable information about potential health concerns within a vulnerable population. 

In addition, we provide a proof-of-concept for an explainable analytical approach that might be used for patient-specific anomaly detection in the eating and drinking habits of PLWD at home and is also applicable to other adverse health events and long-term health conditions. Future work will involve deploying this approach in the Minder platform to raise alerts for tipping points in household behaviour that can then be carer-validated. We would further measure the usefulness of this algorithm qualitatively by collecting feedback on these alerts from the monitoring team, making any necessary adjustments according to this information. We will also work to expand our algorithm to include information from sensors throughout the household.

Providing means to objectively quantify changes in activity patterns will provide a valuable aspect of inspecting the health and well-being of individual PLWD in a meaningful automated way. This strategy can significantly enhance our capacity to augment the provision of personalized dementia care.

\section{Code and Data Availability}
    Code has been made publicly available at: \href{https://github.com/NVFL/Markov-Chain-Model}{https://github.com/NVFL/Markov-Chain-Model}. Anonymized data has been made publicly available at: \href{https://github.com/NVFL/Markov-Chain-Model}{https://github.com/NVFL/Markov-Chain-Model}.
 
\section*{Acknowledgements}
	\setlength{\parindent}{0pt}

We would like to acknowledge support from the UK Dementia Research Institute, Care Research and Technology Centre and Surrey and Borders Partnership NHS Trust. A particularly thank you to Helen Lai for allowing us to use her in-home monitoring system figures which we have used as part of our cohort description and in-home monitoring system figure.

\vspace{0.5cm}

\textbf{Acknowledgement list for UK Dementia Research Institute (UK DRI) Care Research \& Technology (CR\&T) Centre publications using the MINDER core dataset.}

\vspace{0.5cm}

\textbf{Leadership and Management Infrastructure:} 

Centre Director: Professor David Sharp

Co-Director: Professor Payam Barnaghi

Centre Manager: Danielle Wilson

Health and Social Care Lead: Sarah Daniels

Project Managers: Mara Golemme and Zaynab Ismail, Imperial College London

Group Leaders: Professor David Sharp, Professor Payam Barnaghi, Professor Paul Freemont, Dr Ravi Vaidyanathan, Professor Tim Constandinou, Imperial College London 
Professor Derk-Jan Dijk, University of Surrey

\textbf{Groups:} 

\textbf{\emph{Behaviour and Cognition led by Prof David Sharp}}

Michael David, MD, 
Martina Del Giovane,
Neil Graham, MD, PhD,
Naomi Hassim,
Magdalena Kolanko, MD,
Helen Lai,
Lucia Li, MD,
PareshMalhotra, MD, PhD,
Emma Jane Mallas, PhD,
Sanara Raza,
Greg Scott, MD,
Alina-Irina Serban,
Eyal Soreq, PhD,
Tong Wu, PhD

\textbf{\emph{Biosensor Hardware led by Prof Timothy Constandinou}}

Alan Bannon, PhD,
Shlomi Haar, PhD,
Charalambos Hadjipanayi,
Ghena Hammour, 
Bryan Hsieh, 
Adrien Rapeaux, PhD

\textbf{\emph{Robotics and AI Interfaces led by Dr Ravi Vaidyanathan}}

Weiguang Huo, PhD,
Maria Lima,
Maitreyee Wairagkar, PhD
 
\textbf{\emph{Machine Intelligence led by Professor Payam Barnaghi}}

Nan Fletcher-Lloyd, 
Hamed Haddadi, PhD,
Valentinas Janeiko,
Anna Joffe,
Samaneh Kouchaki, PhD,
Viktor Levine,
Honglin	Li,
Francesca Palermo,
Mark Woodbridge,
Yuchen Zhao, PhD,
Alexander Capstick

\textbf{\emph{Point of Care Diagnostics led by Professor Paul Freemont}} 

Loren Cameron, PhD,
Michael Crone, PhD,
Kirsten	Jensen, PhD,
Martin Tran

\textbf{\emph{Sleep and Circadian led by Professor Derk Jan Dijk}}

Ullrich	Bartsch, PhD,
Ciro  Della Monica, PhD,
Kiran GR Kumar, PhD,
Damion Lambert,
Sara Mohammadi Mahvash, PhD,
Vikki Revell, PhD

\textbf{\emph{Helix}}

Matthew Harrison,
Sophie Horrocks,
Lenny Naar

\textbf{Site Investigators and Key Personnel:} 

\textbf{\emph{Surrey and Borders Partnership NHS Foundation Trust (Site and Sponsor)}}

Chief Investigator: Professor Ramin Nilforooshan

Research and Development Manager: Jessica True

Research Co-ordinator: Emily Beale

Clinical Monitoring Team: Vaiva	Zarombaite, Lucy Copps, Olivia Knight

\textbf{\emph{Brook Green Medical Centre }}

Principal Investigator: Dr David Wingfield

\textbf{\emph{Software Engineers, Data Analysts and Technical Staff}}

Severin Skillman, 
Anna Joffe, 
Viktor Levine, 
Valentinas Janeiko,  
Eyal Soreq, 
Helen Lai, 
Martina Del Giovane

\textbf{\emph{Clinical Personnel}}

Michael David, 
Magdalena Anita Kolanko

\section*{Author Contributions}
	Fletcher-Lloyd, N., Serban, A. I., Soreq, E., and Barnaghi, P., conceptualized the study. Fletcher-Lloyd, N. analysed the data, developed the analytical model, and drafted the manuscript. Soreq, E. and Serban, A. I. assisted with data pre-processing and LME modelling. Fletcher-Lloyd, N., Serban, A. I., and Soreq, E. interpreted the results. Barnaghi, P. and Soreq, E. revised the manuscript. 
\section*{Competing Interests}
	The authors declare no competing conflicts of interest.
\section*{Funding}
	This project was supported by the Engineering and Physical Sciences Research Council (EPSRC) PROTECT Project (grant number: EP/W031892/1) and the UK DRI Care Research and Technology Centre funded by the Medical Research Council (MRC) and Alzheimer’s Society (grant number: UKDRI-7002). Alina-Irina Serban was supported by the UK Research and Innovation Centre for Doctoral Training in Artificial Intelligence (UKRI CDT in AI) for Healthcare, see \url{http://ai4health.io} (grant number: P/S023283/1).
 
\bibliographystyle{unsrt}
\bibliography{ref}

	
\end{document}